\pdfoutput=1
\documentclass[journal, onecolumn]{IEEEtran}

\usepackage{amsthm}
\usepackage{amssymb}
\usepackage{amsmath}
\usepackage{amsfonts}
\usepackage{hyperref}
\usepackage{float}
\usepackage{xspace}
\usepackage{algorithm}
\usepackage{algpseudocode}
\usepackage{algorithmicx}
\usepackage{mathrsfs}
\usepackage{cite}
\usepackage{array}
\usepackage{subcaption}
\usepackage{amsmath,amssymb}

\ifx\pdfoutput\undefined
\usepackage{graphicx}
\else
\usepackage[pdftex]{graphicx}
\usepackage{epstopdf}
\fi

\usepackage{ifthen}
\usepackage[usenames, dvipsnames]{color}
\newboolean{showcomments}
\setboolean{showcomments}{true}

\newcommand{\dennice}[1]{  \ifthenelse{\boolean{showcomments}}
{\textcolor{Purple}{(Dennice says:  #1)}}{}}

\newcommand{\souvik}[1]{  \ifthenelse{\boolean{showcomments}}
{\textcolor{red}{(#1)}}{}}

\begin{document}

\title{Locating Power Flow Solution Space Boundaries: A Numerical Polynomial Homotopy Approach }

\author{Souvik~Chandra, Dhagash Mehta, Aranya~Chakrabortty
              \thanks{S. Chandra was with the Department
of Electrical and Computer Engineering, North Carolina State University, Raleigh, NC, USA, when the work in this paper was carried out. He is with Schweitzer Engineering Laboratories, Pullman, WA, USA, e-mail: schandr7@ncsu.edu}
             \thanks{D. Mehta was with the Department of Applied and Computational Mathematics and Statistics, University of Notre Dame, Notre Dame, IN, USA, and Centre for the Subatomic Structure of Matter, Department of Physics, School of Physical Sciences, University of Adelaide, Adelaide, South Australia 5005, Australia, when the work in this paper was carried out. He is with the Systems Department at United Technologies Research Center, East Hartford, CT, USA.}
             \thanks{ A. Chakrabortty is with the Department
of Electrical and Computer Engineering, North Carolina State University, Raleigh, NC, USA, e-mail: achakra2@ncsu.edu.} 
\thanks{Mehta was supported by the NSF through grant NSF-ECCS-1509036 and an Australian Research Council DECRA fellowship no. DE140100867.}
\thanks{Chakrabortty was supported partially through NSF grant ECCS 1509137.}}


\maketitle

\begin{abstract}
The solution space of any set of power flow equations may contain different number of real-valued solutions. The boundaries that separate these regions are referred to as power flow solution space boundaries. Knowledge of these boundaries is important as they provide a measure for voltage stability. Traditionally, continuation based methods have been employed to compute these boundaries on the basis of initial guesses for the solution. However, with rapid growth of renewable energy sources these boundaries will be increasingly affected by variable parameters such as penetration levels, locations of the renewable sources, and voltage set-points, making it difficult to generate an initial guess that can guarantee all feasible solutions for the power flow problem. In this paper we solve this problem by applying a numerical polynomial homotopy based continuation method. The proposed method guarantees to find all solution boundaries within a given parameter space up to a chosen level of discretization, independent of any initial guess. Power system operators can use this computational tool conveniently to plan the penetration levels of renewable sources at different buses. We illustrate the proposed method through simulations on 3-bus and 10-bus power system examples with renewable generation.
%
\end{abstract}


\section{Introduction}
 \label{sec:intro}
Over the past few decades, power system networks in different parts of the world have been going through transformational changes in structure and operation due to integration of large amounts of renewable resources with associated new devices such as power electronic converters, loads, smart transformers, and new transmission lines \cite{wind20pc}. Addition of all of these new devices has tremendously facilitated power system operations and markets, but at the same time has also provoked threats on power system stability \cite{kundur1994power}, especially voltage stability \cite{muljadi2014wind,chi2006voltage,zheng2012impact}. Stimulated by several major voltage collapses, the framework of voltage stability as defined in \cite{van1998voltage} has been extensively studied since the 1990s. Voltage stability can be classified into small disturbance and large disturbance categories. Small disturbance voltage stability refers to the ability of a power system to maintain steady voltage levels following small disturbances experienced through continuous changes in load \cite{freitas2005small}. But with exponentially increasing renewable generation, operational uncertainties and variability not only arise from loads but from generations as well, directly affecting the dynamic performance and voltage fluctuations in the grid. Typically, the transient voltage stability limit of the grid is characterized by a power flow solution space boundary or a loadability boundary at which the Jacobian matrix of the power flow equations is singular \cite{kundur1994power}. System operators choose operating points that are well within the loadability boundary to reduce risk of disruption of service by balancing off deficit powers through reserves. But with high penetration of renewables and their associated intermittency this type of balancing will become more difficult from both operational standpoint and economic viability \cite{bitar2012bringing}. Therefore, it will be critical for operators to have complete information of power flow solution boundaries in terms of the injections levels of the different renewable energy sources. 

In this paper we present a new numerical approach for computing all possible solution boundaries of a power flow problem as a function of a parameter set that models the amount of power injected into the grid from extraneous sources. We do not make any specific distinction between the source of these extraneous injections, meaning that these injections can happen through new renewable installations  such as wind and solar generators as well as through new conventional synchronous generators. Several papers have been written in recent past to address similar problems of finding the power flow solution boundary in a certain parameter space \cite{jepson1985folds,rheinboldt1982computation,hiskens2001exploring}. These methods identify the trajectory of the solution boundaries by continuation-based tracking, starting from an initial guess. The work that is most relevant to our problem is by Hiskens {\it et al.} in \cite{hiskens2001exploring} that proposes a gradient based predictor-corrector method to identify the power flow solution boundary in a given multi-dimensional parameter space. However, the success of this method is critically dependent on two factors, namely, a good initial guess and continuity of the solution boundaries, both of which may be difficult to achieve in face of variability due to intermittent power injection from renewables. For example, the solution boundary may be non-smooth due to inequality constraints in the power flow problem occurring due to generator over-excitation limits or reactive power limits. \cite{karystianos2007maximizing,perninge2011validity}. Moreover, depending on different levels of penetration and their relative locations with respect to the existing generators and loads, the structure of these boundaries may potentially change. Therefore, continuity-based methods such as those in \cite{hiskens2001exploring} may not be able to find all solution space boundaries over a given parameter space. 

In contrast, our approach in this paper is based on a numerical polynomial homotopy-based continuation method (NPHC). This approach guarantees to find all solution boundaries within a parameter space, and can, therefore, help an operator to make a more thorough evaluation of voltage stability limits. The principal distinction of our algorithm from \cite{hiskens2001exploring} is that our method does not look for the singularity of the Jacobian of the power flow equations. Rather we compute all the isolated real solutions of the power flow equations directly at different points on the parameter space. At the points on the solution boundary, which are also denoted as bifurcation points, the number of real-valued solutions of the power flow problem changes. Therefore, one only needs to identify those points in the parameter space. The advantage is that one can obtain all possible equilibria for a given parameter set, not just a local equilibrium that is closest to the initial guess \cite{salam1989parallel,ajjarapu1992,ma1993efficient,liu2005toward}.
NPHC has recently been demonstrated on IEEE prototype power system models in \cite{mehta2014a}. The algorithm in this paper is inspired by a variant of NPHC known as parameter-coefficient NPHC, reported in \cite{li1989cheater,morgan1989coefficient,bates2012paramotopy}. 

The remainder of this paper is organized as follows. In Section II, we introduce the problem statement with the formulation of the power system equilibrium model and a discussion on solving the power flow boundaries via the traditional continuation-based methods as in \cite{hiskens2001exploring}. In Section III we solve the power flow equations using the NPHC approach, and highlight on the use of power system network topology to reduce computation. In Section V we present a discussion on a novel choice of the upper bound for the number of solutions of NPHC utilizing the structure of the power system model. Section V also presents case studies on a 3-bus power system with variable active power injection at one bus, and a 10-bus example with two variable renewable power generators. Section VI concludes the paper with discussions and future directions of research.

\section{Power flow solution boundaries in power systems with renewable generation}
 \label{sec:equilibria_anal}
\subsection{Problem formulation}
 \label{subsec:equi_equation}
We consider a power system with $N$ buses and $n$ generators. These generators are either classified as synchronous generators with indices belonging to a set $\mathcal{S}=:\{1,\dots,n_s\}$, or generators with renewable sources with indices belonging to a set $\mathcal{R}=:\{1,\dots,n_r \}$, where $n_s + n_r=n$. Without loss of generality we classify the buses into 3 sets as follows: Bus $1$ to $n_s$ are the synchronous generator buses whose indices belong to a set $\mathcal{N}_{s}=:\{1, \dots,n_s \}$, bus $n_s+1$ to $n_s +n_r$ are the renewable generation buses whose indices belong to a set $\mathcal{N}_{r}=:\{n_s +1,\dots,n \}$ and the rest of the variables are for load buses with their indices belonging to the set $\mathcal{N}_{l}=:\{n+1, \dots,N\}$. The equilibrium for each of the different buses are obtained by considering power balance with the neighboring buses\cite{chandra2015equilibria}. 
The superscript $e$ for any variable is used to indicate its equilibrium value(s).

Usually in a power flow problem, one of the synchronous generator bus is considered to be the slack bus, which in this case is assumed to be bus $1$ without any loss of generality. Thus the reference voltage level $\left| {{V_{1}^e}} \right|$ and  angle ${{\theta_{1}^e}}$ are respectively equal to 1 and 0. Correspondingly, the steady-state active and reactive power flow, ${P_{1}^e}$ and ${Q_{1}^e}$ from slack bus $1$ can be expressed as,
 \begin{subequations}
 \label{eqn:power_slbus}
\begin{align}
 {P_{1}^e} =&{\rm{Re}}\left\{
{\sum\limits_{k =2}^N { {{\left(
{\frac{{\left(1-V_{k}^e \right)}}{{{Z_{1k}}}}} \right)}^*}} } \right\} + P_{L1}^e, \\
 {Q_{1}^e}=& {\mathop{\rm
Im}\nolimits} \left\{ {\sum\limits_{k = 2}^N
{{{\left( {\frac{{{\left(1-V_{k}^e \right)}}}{{{Z_{1k}} }}} \right)}^*}} }
\right\} + Q_{L1}^e ,
\end{align}
\end{subequations}
where $k$ denotes the index of neighboring buses, $V_{k}$ is the voltage of the bus $k$ and $Z_{1k}$ is the line impedance connecting buses $1$ and $k$. $P_{L1}$ and $Q_{L1}$ are the active and reactive power of the loads connected to bus $1$. All other synchronous generator buses, $i\in\mathcal{N}_{s}$, such that $i\neq 1 $ are considered to be PV buses for which the active power  ${P_{i}^e}$  and $\left| {{V_{i}^e}} \right|$ are specified. The power balance for bus $i$ with the neighboring buses is shown as below, 
\begin{subequations}
\label{eqn:power_syncbus}
\begin{align}
 {P_{i}^e}=& {\rm{Re}}\left\{
{\sum\limits_{k = 1,k \ne i}^N { {{V_{i}^e}}{{\left(
{\frac{{{V_{i} ^e}-V_{k}^e}}{{{Z_{ik}}}}} \right)}^*}} } \right\} + P_{Li}^e \\
 {Q_{i}^e} =& {\mathop{\rm
Im}\nolimits} \left\{ {\sum\limits_{k = 1,k \ne i}^N
V_{i}^e{{{\left( {\frac{{{V_{i}^e}-V_{k}^e}}{{{Z_{ik}} }}} \right)}^*}} }
\right\} + Q_{Li}^e .
\end{align}
\end{subequations}
The bus $j\in\mathcal{N}_{r}$, connected to a renewable source, may be either PV or PQ depending on the type of the renewable generators. But in this work, for the sake of simplicity we limit our model to only those types of renewable generators for which internal control of both active and reactive power at the output bus is available. Thus, we consider the renewable bus to be a PV bus which maintains a given reference voltage $\left| {{V_{j}^e}} \right|$ and injects active power  ${P_{j}^e}$. The power balance is shown as below,
\begin{subequations}
\label{eqn:power_windbus}
\begin{align}
 {P_{j}^e}=& {\rm{Re}}\left\{
{\sum\limits_{k = 1,k \ne j}^N { {{V_{j}^e}}{{\left(
{\frac{{{V_{j} ^e}-V_{k}^e}}{{{Z_{jk}}}}} \right)}^*}} } \right\} + P_{Lj}^e \\
 {Q_{j}^e} =& {\mathop{\rm
Im}\nolimits} \left\{ {\sum\limits_{k = 1,k \ne j}^N
V_{j}^e{{{\left( {\frac{{{V_{j}^e}-V_{k}^e}}{{{Z_{jk}} }}} \right)}^*}} }
\right\} + Q_{Lj}^e .
\end{align}
\end{subequations}
The total active and reactive power injections at these buses are assumed to be a product of a factor $\lambda_j$ and constant unit powers $P_{r}^{e}$ and $Q_{r}^{e}$ respectively, where $j \in \mathcal{N}_r$. Here, $\lambda_j$, for example, may represent the total number of wind turbines behind the point of common coupling with the grid, assuming every turbine produces the same amount of power in equilibrium. In general, $\lambda_j$ is an indicator of the total amount of injection at bus $j$, and will be treated as a variable in the power flow equations. The variable parameter $\lambda_j$ can also be incorporated in \eqref{eqn:power_syncbus} for an increase in synchronous generation resulting in transmission expansion. For convenience of analysis we limit the use of $\lambda_j$ to \eqref{eqn:power_windbus} to introduce variability only through the renewable sources.  
Correspondingly, equation \eqref{eqn:power_windbus} can be rewritten as,
\begin{subequations}
\label{eqn:power_windbus1}
\begin{align}
 {\lambda_{j}}{P_{r}^e}=& {\rm{Re}}\left\{
{\sum\limits_{k = 1,k \ne j}^N { {{V_{j}^e}}{{\left(
{\frac{{{V_{j} ^e}-V_{k}^e}}{{{Z_{jk}}}}} \right)}^*}} } \right\} + P_{Lj}^e \\
 {\lambda_{j}}{P_{r}^e} =& {\mathop{\rm
Im}\nolimits} \left\{ {\sum\limits_{k = 1,k \ne j}^N
V_{j}^e{{{\left( {\frac{{{V_{j}^e}-V_{k}^e}}{{{Z_{jk}} }}} \right)}^*}} }
\right\} + Q_{Lj}^e ,
\end{align}
\end{subequations}
For load bus, $j\in\mathcal{N}_{l}$, usually the active and reactive load power ${P_{Lj}^e}$, ${Q_{Lj}^e}$  are specified. The power balance is shown as below, 
\begin{subequations}
\label{eqn:power_loadbus}
\begin{align}
 0=& {\rm{Re}}\left\{
{\sum\limits_{k = 1,k \ne j}^N { {{V_{j}^e}}{{\left(
{\frac{{{V_{j} ^e}--V_{k}^e}}{{{Z_{jk}}}}} \right)}^*}} } \right\} + P_{Lj}^e \\
 0 =& {\mathop{\rm
Im}\nolimits} \left\{ {\sum\limits_{k = 1,k \ne j}^N
V_{j}^e{{{\left( {\frac{{{V_{j}^e}-V_{k}^e}}{{{Z_{jk}} }}} \right)}^*}} }
\right\} + Q_{Lj}^e .
\end{align}
\end{subequations}
 The complete set of power flow equations for a power system with renewable injections is given by the set of nonlinear algebraic equations shown in \eqref{eqn:power_slbus}-\eqref{eqn:power_loadbus}. These equations are parametrized by $\lambda \in \mathbf{R}^{n_r}$, the vector of all renewable penetration levels. \eqref{eqn:power_slbus}-\eqref{eqn:power_loadbus} can be represented as a set of equations $P$ as shown below,
 \begin{equation}
\label{eqn:power_flow1}
 P(x,\lambda ) = 0.
\end{equation}
 where $x \in \mathbf{R}^{2N}$ is a vector of unknown power flow variables. Equation \eqref{eqn:power_flow1} for the slack bus is solved for $x$ which includes,  ${P_{1}^e}$ and ${Q_{1}^e}$ for slack bus, ${Q_{i}^e}$ and ${{\theta_{i}^e}}$ where $i \in (\mathcal{N}_{s}\cup \mathcal{N}_{r}), i\neq 1$ for the PV buses, ${{\theta_{j}^e}}$ and  $\left|{V_{j}^e}\right|$ where $j \in \mathcal{N}_{l}$ for the PQ buses. 
  
\subsection{Problem to be solved}
 \label{sec:problem_form}
 The power flow solution boundary represents those solutions of the power flow equations \eqref{eqn:power_slbus}-\eqref{eqn:power_loadbus} 
 for which the Jacobian of the equations is singular. From \eqref{eqn:power_flow1}, it is clear that the Jacobian is a function of renewable penetration levels $\lambda$.
 The objective of this work is to find out all boundary solutions of \eqref{eqn:power_slbus}-\eqref{eqn:power_loadbus} for a given set of points in the parameter space of $\lambda$, and obtain the corresponding solution boundaries. 
\subsection{The traditional method}
 \label{sec:cont_method}
 In \cite{hiskens2001exploring} Hiskens {\it et al.} propose a technique based on a predictor-corrector method to identify the power flow solution boundary of $P$ defined in \eqref{eqn:power_flow1} over a parameter space formed by $\lambda$. The solution method consists of two steps. Initially the solution space boundaries are represented as a solution of the set of equations as shown below,
 \begin{subequations}
 \label{eqn:hiskens1}
\begin{align}
 P(x,\lambda ) &= 0\\
g(x,\lambda ,v) &= {P_x}(x,\lambda).v = 0\\
h(v) &= {v^T}v = 1.
\end{align}
\end{subequations}
Here the Jacobian matrix of $P$ with respect to $x$ is given by $P_x$. $v\in\mathbf{R}^{2N}$ is a right 
eigenvector corresponding to a singular eigenvalue of $P_{x}$. As a first step of the method, one of the parameters, say $\lambda_{1}$ is varied while all others are kept constant. Consequently, \eqref{eqn:hiskens1} turns into a set of $2n+1$ equations with $2n+1$ unknowns which can be solved via computational methods such as Newton-Raphson with an initial guess. The solution of this first step, say $x_{0}$, will provide an initial point on the solution space boundary. Starting from $x_{0}$, a gradient based predictor-corrector tracking is implemented to obtain the other points on the boundary. First, the algebraic equations in \eqref{eqn:hiskens1} are represented as,
\begin{equation}
\label{eqn:hiskens3}\phi (z) = \left[ {\begin{array}{*{20}{c}}
{P(z)}\\
{g(z)}\\
{h(z)}
\end{array}} \right],
\end{equation}
where $z =\left[{\begin{array}{*{20}{c}}x&v&\lambda\end{array}}\right]^{'}.$ An initial guess $z_{0}$ is obtained which satisfies the condition $\phi (z_0)=0.$ Starting from $z_{0}$,
the algorithm iteratively solves for each $z_{i}$ by solving the following equations at each step $i$,
\begin{subequations}
 \label{eqn:hiskens4}
\begin{align}
\phi (z_i) &= 0,\\
{\left(z_i-z_{i-1}\right)}^{'}\nu &= \epsilon.
\end{align}
\end{subequations}
$\epsilon$ is a small constant that determines the accuracy of the numerical approximation. $\nu$, on the other hand, is a vector that is tangential to the solution boundary at
$z_{i-1}.$ Repeated solutions of $z_{i}$ can provide closed boundary contours of the power flow equations. Since this method tracks the solution boundary iteratively, the choice of the initial point $z_{0}$ is particularly important for the success of this method. As specified earlier, the boundaries can be often disjoint and non-smooth. In such scenarios this method does not guarantee to obtain all solution boundaries. To solve this problem, we introduce a numerical homotopy continuation method, detailed in the next section.  
\section{Parameter Homotopy Continuation Algorithm}
\label{sec:numer_homotop}
In this section we solve for {\it all} real equilibria of the power system model shown in \eqref{eqn:power_slbus}-\eqref{eqn:power_loadbus} using the parameter-coefficient NPHC method \cite{li2003solving,sommese2005numerical}. As a preconditioning, we represent the algebraic equations shown in \eqref{eqn:power_slbus}-\eqref{eqn:power_loadbus} as multivariate polynomials. To achieve this, the voltage phasor $V_{j}^{e}$ at any bus $j\in N$ is expanded in terms of its magnitude $\left|{V_{j}^e}\right|$ and angle ${{\theta_{j}^e}}$, such that $V_{j}^{e}=\left|{V_{j}^e}\right|\cos{{{\theta_{j}^e}}}+\mathbb{J}\left|{V_{j}^e}\right|\sin{{{\theta_{j}^e}}}$, $\mathbb{J}$ being the complex operator. Using such expansion, energy balance in \eqref{eqn:power_windbus} for instance can be expressed as,
\begin{subequations}
\label{eqn:power_windbus_mod}
\begin{align}
P_j^e = &{\rm{ Re}}\left\{ {\sum\limits_{k = 1,k \ne j}^N {{S_{jk}}} } \right\} + P_{Lj}^e\\
 Q_j^e = &{\rm{ Im}}\left\{ {\sum\limits_{k = 1,k \ne j}^N {{S_{jk}}} } \right\} + Q_{Lj}^e ,
\end{align}
\end{subequations}
where power flow between bus $j$ and $k$ can be represented as,
\footnotesize
\[{S_{jk}} = \frac{{\left| {V_j^e} \right|\left( {\left| {V_j^e} \right|{{\cos }^2}\theta _j^e - \left| {V_k^e} \right|\left( {\cos \theta _j^e\sin \theta _k^e - \mathbb{J}\sin \left( {\theta _j^e - \theta _k^e} \right)} \right)} \right)}}{{{\rm{arg}}\left( {{Z_{jk}^*}} \right)}}.\]
\normalsize

${Z_{jk}^*}$ represents the complex conjugate of the line impedance connecting bus $j$ and bus $k$. In the modified system of equations, power balance of slack bus 1 has $P_1^e$ and $Q_1^e$ as variables yielding two first-order equations. The power balance of PQ bus $j$ have $\left|{V_{j}^e}\right|\cos{{{\theta_{j}^e}}}$ and $\left|{V_{j}^e}\right|\sin{{{\theta_{j}^e}}}$ as the variables with two second-order equations. 
Both equations are in quadratic polynomial forms, as indicated in \eqref{eqn:power_windbus_mod}. For a PV bus, however, the variables are $Q_j ^e$ and $\theta_j^e$, due to which \eqref{eqn:power_windbus_mod} no longer retains its polynomial form for this type of a bus. To solve this problem, the unknowns are alternatively defined as $Q_j ^e$, $\cos\theta_j^e$, and $\sin \theta_j^e$,  all of which are independent of each other. An extra constraint equation now needs to be added as
\begin{equation}
\cos{{{\theta_{j}^e}}}^{2}+\sin{{{\theta_{j}^e}}}^{2}=1,
\label{eqn:sin_cos_rel}
\end{equation}
where $j \in \mathcal N_s \cup \mathcal N_r$. As a consequence, $n-1$ more equations are added for the $n-1$ PV buses considered in our system. The resulting system turns into a set of $2N+n-1$ quadratic equations in $x$ and $\lambda$ as,
\begin{equation}
\mathbb{P}(x,\lambda)=0,
\label{eqn:poly_struc}
\end{equation}
where, $x \in \mathbf{R}^{(2N+n-1)}$ is the vector of unknowns, and $\lambda \in \mathbf{R}^{n_r}$ is a constant parameter vector. It is noted that the symbol for the function $P$ in (6) has been changed to $\mathbb P$ to indicate that the latter represents a set of quadratic equations in the variable $x$. In essence, both equations (6) and (12) are equivalent.

As a first step of the NPHC method, an upper bound of the number of complex isolated solution of \eqref{eqn:poly_struc} is determined. Next, a homotopy $H(x,t)$ is defined as shown below,
\begin{equation}
H(x,t)=\eta_{h}(1-t)\mathbb{Q}(x)+t\; \mathbb{P}(x),
\end{equation}
where $\mathbb{Q}(x)$ is an arbitrary start system which is easily solvable. $\mathbb{Q}(x)$ is chosen in such a way that the number of isolated solutions of $\mathbb{Q}(x)=0$ is equal to the estimated upper bound of isolated solutions of \eqref{eqn:poly_struc}. $\eta_h$ is a generic complex number and $t$ is a continuous parameter varying from 0 to 1. Therefore, the solution set of $H(x,t)=0$ for $0\le t\le 1$ actually consists of a finite number of smooth paths parametrized by $t\in[0,1)$.  For a generic $\eta_h\in\mathbb{C}$, 
it is proven in \cite{sommese2005numerical} that each of the paths will be well-behaved, i.e., either they will converge to $H(x,1) = 0$, or will diverge to infinity. Hence, for a generic value of $\eta_h$, the NPHC method guarantees to find all isolated complex solutions \cite{morgan1987computing} of  $\mathbb{P}(x)=0$. The crux of the algorithm is to track each solution of $H(x,t)=0$ for $t\in[0,1)$ using an efficient predictor-corrector method \cite{bates2013numerically} to obtain all complex solutions for  $\mathbb{P}(x)=0$. Next we discuss the issue of maximum number of paths which should be tracked for a system of polynomials $\mathbb{P}(x)$ to guarantee  all complex solutions of $\mathbb{P}(x)=0$.

Now, a system of $m$ polynomials can have a maximum of $\prod_{i=1}^{m}{d_{i}}$ number of isolated complex solutions, where $d_{i}$ is the degree of the $i^{th}$ polynomial, as specified by the classical B\'ezout theorem. This poses an upper bound on the number of solutions to be tracked for NPHC method known as classical B\'ezout bound (CBB). For solving our power flow problem of \eqref{eqn:poly_struc} there will be ${(2N+n-1)}$ algebraic equations, each of degree 2. Thus the number of paths to be tracked is $2^{(2N+n-1)}$. This translates to the fact that in our model the number of paths to be tracked grows exponentially with the number of buses. 
Converting the power flow equations to their quadratic forms as in \eqref{eqn:poly_struc} actually pays off here by allowing a tighter upper bound on the number of complex isolated solutions or the CBB. However, this crude upper bound does not capture the specific complex algebraic structure of the polynomials of \eqref{eqn:poly_struc}. Moreover, we are interested in solving \eqref{eqn:poly_struc} over a parameter space given by specific penetration levels $\lambda$. Solving such parametric systems for every parameter-point from scratch using the NPHC method can be computationally very expensive. We, therefore, use a more sophisticated approach based on a variant of NPHC, parameter-coefficient homotopy \cite{morgan1989coefficient},
an earlier version of which was called Cheater's homotopy \cite{li1989cheater}. This method uses the fact that for a parametric system of polynomial equations, the maximum number of isolated complex solutions over all parameter-points is same for a generic complex parameter-point. 
 \begin{figure*}[t]
        \begin{subfigure}[b]{0.45\textwidth}
                \centering
                \includegraphics[width=\textwidth]{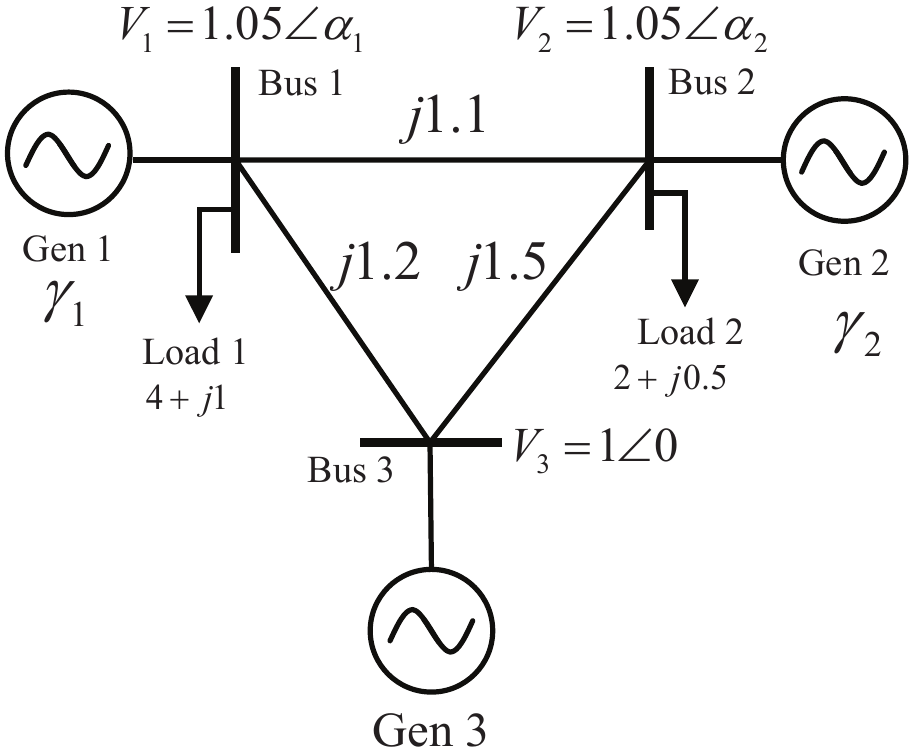}
                \caption{Power system model}
               \label{fig:3_bus_power_sys}
        \end{subfigure}%
        ~~~ 
             \begin{subfigure}[b]{0.45\textwidth}
                \centering
                \includegraphics[width=\textwidth]{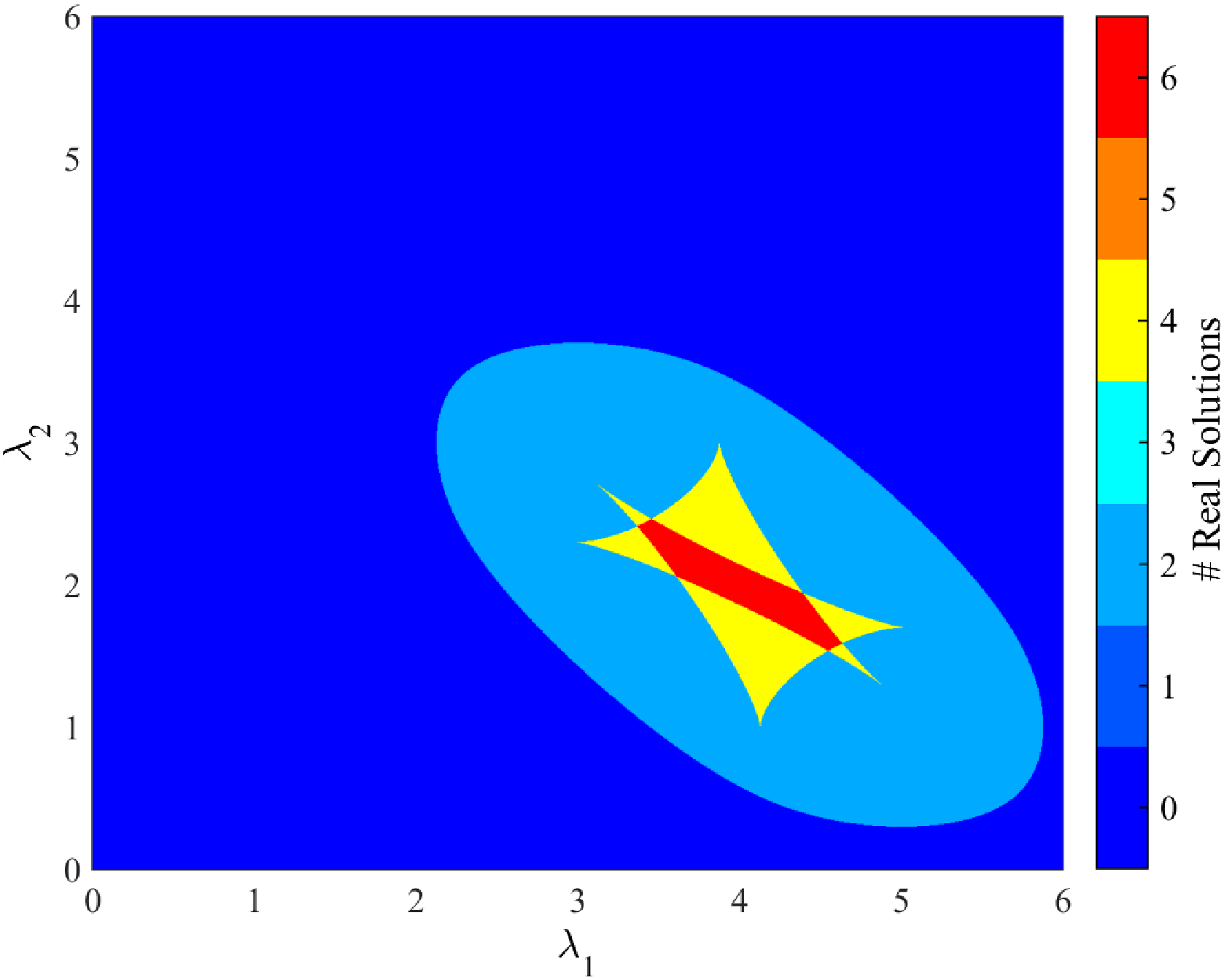}
                \caption{Number of real solutions of the power flow problem}
                \label{fig:3_bus_souvik_sol}
        \end{subfigure}
        \caption{ 3-bus power system with two parameters}
\end{figure*}
Hence, we can solve $\mathbb{P}(x, \lambda)=0$ at a \textit{generic} complex parameter-point $\lambda^{*} \in \mathbb{C}^{m}$, using the NPHC method with the help of some crude upper bound on the number of complex solutions such as the CBB. Although such a complex parameter-point is physically not meaningful as $\lambda$ represents the level of renewable penetration, solving the system at such a point reduces the computation for all other physically relevant parameter-points.
Next, we choose $\mathbb{P}(x, \lambda^{*})=0$ as the start system for all other parameter-points $\lambda \in \mathbb{C}^{m} - \{ \lambda^{*} \}$. 
Each solution of this start system needs to be tracked with the following homotopy:
\begin{equation}
\label{eqn:homotopy}
H(x,\lambda, t)= t\; \mathbb{P}(x, \lambda^{*}) + (1 - t)\; \mathbb{P}(x, \lambda)=0,
\end{equation}
from $t=1$ to $t=0$. This procedure again guarantees {\it all} isolated complex solutions at each of the chosen parameter-points, independent of the upper bound chosen to solve the system at $\lambda^{*}$ in the first step. In our simulations we will show that depending on operating conditions and structure of the power system, the number of paths to be tracked in the first step (e.g. the CBB) for power system models can dramatically reduce to a very small integer in the second step. The method thus becomes computationally very cheap once the first step is solved.
Moreover, the process of solving the second step for a parameter vector $\lambda^1$ is independent of a different parameter vector $\lambda^2$. Hence, if one intends to solve \eqref{eqn:poly_struc} for multiple $\lambda$ at the same time the process can be parallelized.
For our simulations, we used a novel computational package called \emph{Paramotopy} \cite{bates2012paramotopy} which efficiently implements the above mentioned procedure with appropriate parallelization. Once {\it all} real solutions are obtained at each parameter point one can accurately estimate the power flow solution boundaries where 
 the number of real solutions changes, given the parameter space is densely represented. 
 \subsection{On the Network Topology and Upper Bound on the Number of Equilibria}
An actual system of polynomials may have fewer complex solutions as compared to its CBB. Thus, in the parameter-coefficient homotopy, it would be computationally wasteful to track the paths which would eventually diverge. 
Thus, for solving large sets of power flow equations one should exploit the underlying structure or the sparsity of the network connectivity of any given power system model, and try to compute a tighter upper bound. Recent reviews on the existing results on upper bounds are provided in \cite{molzahn2015toward} and \cite{mehta2015recent}. An upper bound of $\binom{2N - 2}{N-1}$ was computed in \cite{baillieul1982,li1987numerical,marecek2014power} for a generic power flow problem with $N$ buses, although it did not still exploit the network topologies. In \cite{guo1990}, the number of complex solutions for networks with cliques with exactly one common node was shown to be equal to the product of number of complex solutions for the individual cliques as independent networks. In \cite{molzahn2015toward},
this result is extended to other related network topologies, though several of the patterns for the particular topologies are still not well understood. 

Sparsity of network connectivity, for example, may result in lesser number of complex solutions of (12) than the usual CBB which is $2^{(2N+n-1)}.$
This is because the coefficients of the polynomial are related to each other according to the network topology, and hence certain sparsity patterns may be such that a large subset of all the possible monomials of degree up to the highest degree of the polynomials do not appear in those polynomials. For example, consider the power balance of bus 1 in the 3-bus system in Figure \ref{fig:3_bus_power_sys} which can be written as,
\begin{subequations}
\label{eqn:power_3bus_bus1}
\begin{align}
P_1^e = & {\rm{Re}}\left\{ {\sum\limits_{k = 2}^3 {{S_{1k}}} } \right\} + P_{L1}^e \\
Q_1^e = & {\rm{Im}}\left\{ {\sum\limits_{k = 2}^3 {{S_{1k}}} } \right\} + Q_{L1}^e .
\end{align}
\end{subequations}
where the power flow from bus 1 to any bus $k$ is given as,
\small
\[{S_{1k}} = \frac{{\left| {V_1^e} \right|\left( {\left| {V_1^e} \right|{{\cos }^2}\theta _1^e - \left| {V_k^e} \right|\left( {\cos \theta _1^e\sin \theta _k^e - \mathbb{J}\sin \left( {\theta _1^e - \theta _k^e} \right)} \right)} \right)}}{{{\rm{arg}}\left( {{Z_{1k}^*}} \right)}}.\]
\normalsize

The quadratic equations in \eqref{eqn:power_3bus_bus1} are not the densest polynomials of degree $2$ as the interaction of the unknown variables occur in only a specified structural form. For example, if the physical connection or the transmission line between bus 1 and bus $k$ does not exist,  then $1/\rm{arg}\left({{\it Z}_{1k}^*}\right)=0$. This eliminates the interaction terms of the variables associated with bus 1 and bus $k$. Thus the number of complex solutions at a generic complex parameter point can always be expected to be lesser than the CBB, particularly for a power flow problem.

\section{Examples }
\label{sec:example}
\subsection{3 bus system}
\label{3 bus system}
\begin{figure*}[t]
        \begin{subfigure}[b]{0.52\textwidth}
               \centering
                \includegraphics[width=\textwidth]{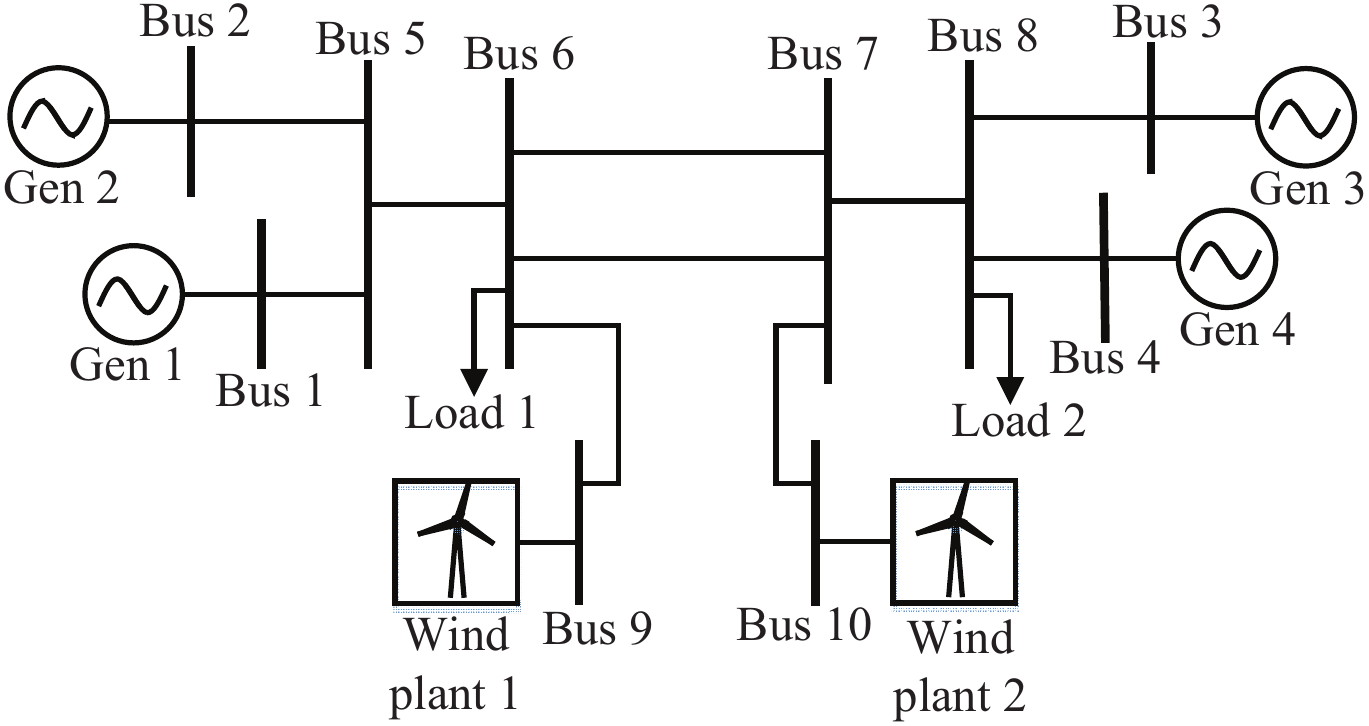}
                \caption{power system model}
                \label{fig:10_bus_power_sys}
        \end{subfigure}
        ~~
             \begin{subfigure}[b]{0.42\textwidth}
               \centering
                \includegraphics[width=\textwidth]{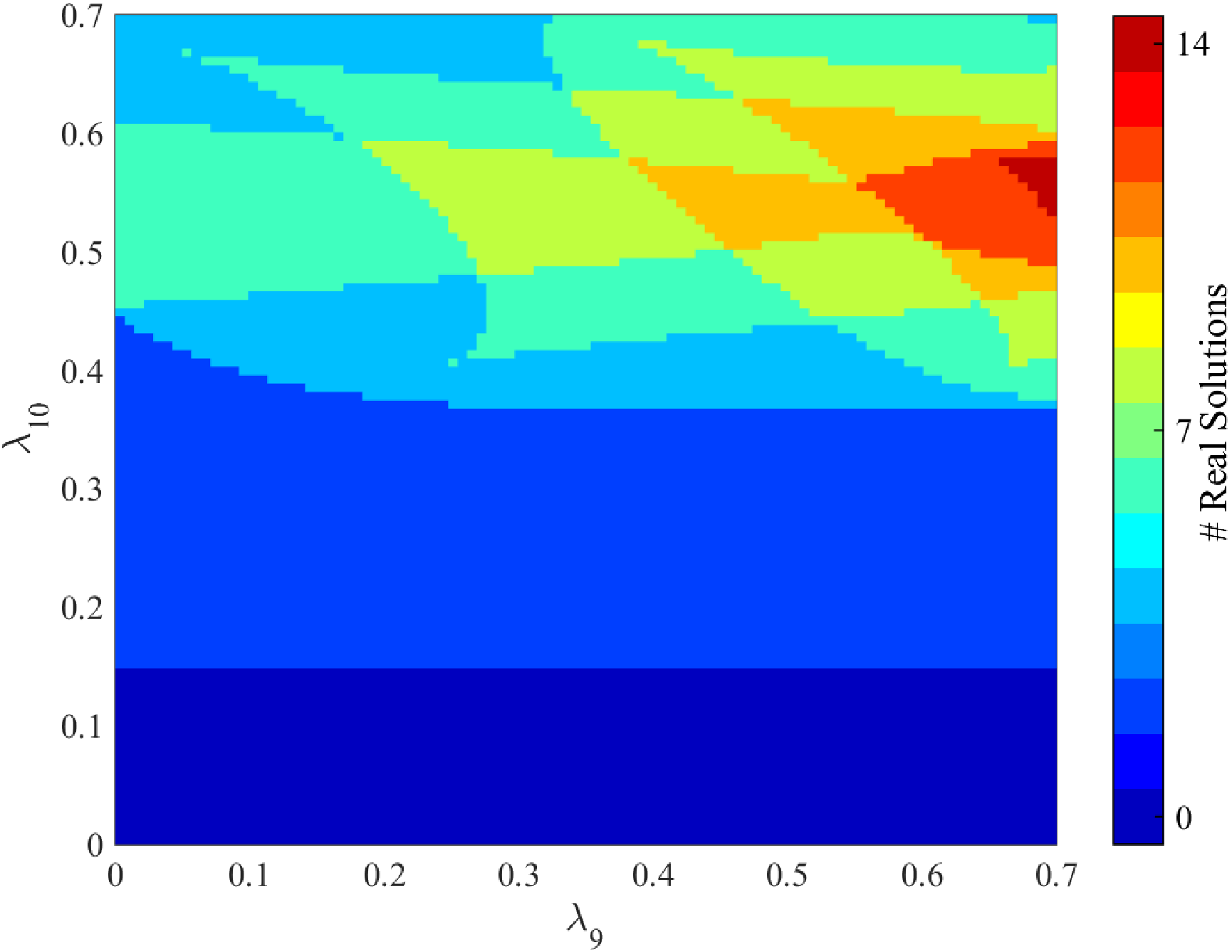}
                \caption{Number of real solutions of the power flow problem}
                \label{fig:10_bus_souvik_sol}
        \end{subfigure}
        \caption{ 10-bus power system with two wind power plants with penetration levels $\lambda_{9}$ and $\lambda_{10}$ respectively}
\end{figure*}
\begin{figure}[t]
                \centering
               \includegraphics[width=0.45\textwidth]{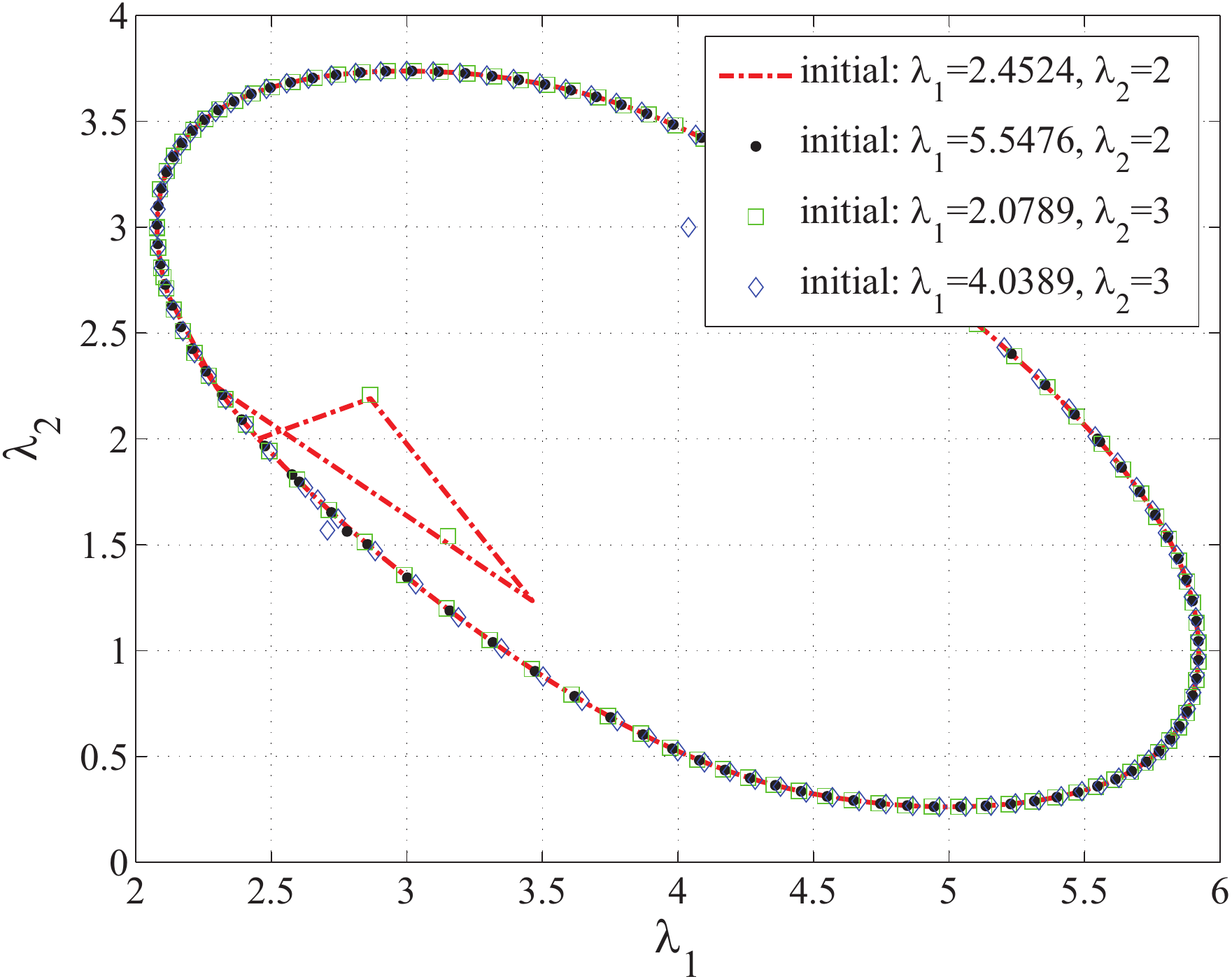}
                \caption{Power flow solution boundary tracking with various initial points}
                \label{fig:3_bus_hisken's_sol}
        \end{figure}
First we explore the solution space boundary for a 3-bus power system as shown in Figure \ref{fig:3_bus_power_sys}, which is a modification 
of an example in \cite{hiskens2001exploring} with added loads. The active power input at bus 1 and 2 are the variable parameters $\lambda_{1}$ and $\lambda_{2}$ respectively. Bus 3 is assumed to be the swing bus whose voltage equals $1\angle 0$. The unknown variables of the power flow problem which constitute the vector $x$ in \eqref{eqn:hiskens1} are the active and reactive power input at bus 3, the reactive powers and angles of bus 1 and 2. 
However as shown in Section \ref{sec:numer_homotop} we represent the angles in rectangular form to limit the order of the algebraic equations to two. Thus we have the sine and cosine of the angle of bus 1 and bus 2 as the unknown variables.
The problem, therefore, reduces to the computation of the unknown vector,

\footnotesize
\[x=\left[ {\begin{array}{*{20}{c}}
{{{P}_3}}&{{Q_3}}&{{Q_1}}&{{Q_2}}&{\sin {\delta _1}}&{\cos {\delta _1}}&{\sin {\delta _2}}&{\cos {\delta _2}}\end{array}} \right]^{'},\]
\normalsize
over a set of parameter values $\lambda_{1}$ and $\lambda_{2}$.
As mentioned earlier in Section \ref{sec:numer_homotop}, we solve the system of equations $\mathbb{P}(x,\lambda^{*})=0$ at a generic complex parameter point $\lambda^{*}$. The generic complex vector has two elements which are chosen from uniform distributions such as $\{a+ib : -1\le a,b \le 1\}$, and are normalized to ensure that they are inside unit circle. For the start problem  $\mathbb{P}(x,\lambda^{*})=0$, we followed the CBB and tracked $2^6=64$ paths to obtain all the isolated complex solutions. However, the start problem for this case yielded only $6$ complex solutions. Correspondingly, in the second stage of the algorithm, following the parameter-coefficient NPHC method 
we to track the 6 paths for each parameter point starting from the start solution. Once all the complex solutions have been obtained the real solutions are identified by doing a numerical check. As seen in Figure \ref{fig:3_bus_souvik_sol} the number of real solutions vary from 0 to 6. 
The boundaries can be identified by the change in the number of real solutions where the Jacobian becomes singular. $(\lambda_1,\, \lambda_2)$ parameter space, as shown in Figure \ref{fig:3_bus_souvik_sol} is discretized into a grid of equispaced parameter points of dimension $100\times 100$. All solutions for each of the discrete points on the parameter space are obtained henceforth by the application of numerical homotopy.


Following Hiskens {\it et al.}\cite{hiskens2001exploring}, as a first step we keep $\lambda_{2}$ fixed, and find an initial point on the solution boundary. Essentially we solve \eqref{eqn:hiskens1} by a Newton-Raphson method with different initial points for $\lambda_{1}.$ When $\lambda_{2}=2$, we find initial points with $\lambda_{1}=2.4534$ and $5.5476$. When $\lambda_{2}=3$, the initial points are $\lambda_{1}=2.0789$ and $4.0389$. As seen in Figure \ref{fig:3_bus_souvik_sol}, all of these points are located on the outer elliptical solution boundary. Correspondingly, in the second step when we solve  \eqref{eqn:hiskens4} in an iterative form starting from these points they only track the outer boundary as shown in Figure \ref{fig:3_bus_hisken's_sol}. Thus following the method in \cite{hiskens2001exploring}, it is difficult to identify all the power flow solution boundaries unless one has a sound knowledge of the solution space for a given set of parameters. On the contrary our alternate numerical homotopy based method can guarantee to identify {\it all} solution boundaries for a given power flow problem. In Figure \ref{fig:3_bus_souvik_sol}, it can be seen that via homotopy continuation method, we could identify regions on the parameter space with different number of real solutions identified by the different colors. The boundaries between these regions are the power flow solution boundaries. It can be noted that our method identifies boundaries between $0,2,4$ and 6 real solutions while the conventional method identifies the boundary between 0 and 2 solutions only. 
\subsection{10 bus system}
\label{10 bus system}
In the next example we use a 10-bus, 4-synchronous machine power system as shown in Figure \ref{fig:10_bus_power_sys}. Two wind power plants are connected at bus 9 and bus 10 whose penetration levels are represented as $\lambda_{9}$ and $\lambda_{10}$ respectively. The parameters of the simulation are given in Appendix \ref{app_10bus}. We vary the active power output of each of the wind plants between 0.1 and 0.7 p.u. on a 100 MVA base and find the solution of the power flow problem for each point on the plane defined by $\lambda_{9}$ and $\lambda_{10}$. We first solve the problem at a generic complex parameter point using the homotopy based algorithm of section \ref{sec:numer_homotop}. The problem has 25 unknown equations which require $2^{25}$ paths to be tracked by continuation in the start problem to ensure all the isolated roots. However, it turns out that the start system has only $2^9$ isolated roots. Correspondingly, we look for only $2^9$ paths in the subsequent steps saving a lot of computational effort in finding roots for actual parameter values. 
Figure \ref{fig:10_bus_souvik_sol} shows the number of real solutions of the power flow problem for given values of $\lambda_{9}$ and $\lambda_{10}$. The different colored regions of Figure \ref{fig:10_bus_souvik_sol} demonstrate varying number of real solutions of the power flow. Thus the boundaries between the regions constitute the power flow solution boundaries. It can be observed that the geometry of the solution boundary for the 10-bus case with varying wind penetration levels is strikingly different as compared to the 3-bus case. Also the number of isolated boundaries are more than that of the 3-bus case. All these observations point to the fact that identifying these boundaries by an initial guess and local approximation is totally intractable for systems with large dimension.
Thus, if the system has multiple wind power plants, then our algorithm can provide the power system operator to choose an optimal set of power injections at the renewable buses. 
The operating points can be post processed for different robustness criteria and placed at a suitable distance away from the loadability boundary of the system. In this case as well, the parameter space $(\lambda_1, \,\lambda_2)$  as shown in Figure \ref{fig:10_bus_souvik_sol} is discretized into a grid of equispaced parameter points of dimension $100\times 100$.

Although finding a novel upper bound on the number of power flow solutions is not our goal, we still observed from the above simulations that the number of complex solutions at a generic complex parameter-point is dramatically small compared to the CBB of the system. It is evident from the simulation results that this number is a new and tighter upper bound on the number of complex isolated solutions compared to the previously known upper bounds. For example, for the 3 bus case, the CBB was $64$ whereas the number of complex solutions at a generic point was only $6$, which is the same as the binomial bound mentioned above. However, for the 10 bus case, the CBB is $2^{23}$ while the number of complex solutions is $2^9 = 512$, which is much smaller compared to the binomial bound $48620$.

\section{Conclusion}
\label{sec:conclu}
We developed a numerical power flow solution method that guarantees to identify
all power flow solution boundaries in a power system in
presence of variable generation parameters. Determination of all solution
boundaries will be critical in the foreseeable future due to the variability imposed
by the rapid intrusion of renewable energy penetration. The essence of our algorithm is based on a homotopy continuation concept, which also has the potential for accommodating topological information of the system. Our future research direction would include deriving an  explicit relationship between the number of solutions and the structure of the power system, that may lead to real-time usage of this tool during different contingencies. 
\section{Acknowledgement}
The authors would like to thank Dr. Daniel K. Molzahn and Dr. Konstantin Turitsyn for their helpful suggestions and discussions on this topic.

\appendices
\section{Model Parameters for the 10-bus system}
\label{app_10bus}
The conventional generators have the following power output on 100-MVA base: 

$P_{1}^{e}=35.91$ p.u., $P_{2}^{e}=17.85$ p.u., $P_{3}^{e}=10.00$ p.u., $P_{4}^{e}=40.00$ p.u.

Line parameters in per unit on 100-MVA base:
\small
$Z_{15}=\left(0.25+j2.50\right)e-3$, $Z_{25}=\left(0.25+j2.50\right)e-3$, $Z_{38}=\left(0.25+j2.50\right)e-3$, $Z_{48}=\left(0.25+j2.50\right)e-3$, $Z_{56}=\left(0.10+j1.00\right)e-3$, $Z_{78}=\left(0.10+j1.00\right)e-3$, $Z_{67}=\left(2.20+j22.00\right)e-3$, $Z_{69}=\left(0.25+j2.50\right)e-3$,
$Z_{710}=\left(0.25+j2.50\right)e-3$
\normalsize

Load parameters in per unit on 100-MVA base: $S_{L6}=30+j9.11$ and  $S_{L8}=70+j20$.

The rated power output of the wind plants at bus 9 and bus 10 are
$P_{r9}^{e}=25$ p.u. and $P_{r10}^{e}=15$ p.u..

\bibliographystyle{unsrt}
\bibliography{wind_biblio}





\end{document}